\documentclass[useAMS,usenatbib,psfig]{mn2e}
\usepackage{graphicx,rotating,times}
\usepackage{longtable,supertabular,multirow}
\usepackage{amsmath,amssymb,rotating}
\usepackage{lscape}
\newcommand{\hii}{{H{\scriptsize II} }}

\newcommand{\nhthree}{\mbox{NH$_3$}}
\newcommand{\nhone}{\mbox{NH$_3$(1,1)}}

\newcommand       \be           {\begin{equation}}
\newcommand       \ee           {\end{equation}}
\newcommand       \yr		{\,{\rm yr }}
\newcommand       \s		{\,{\rm s }}
\newcommand       \pc		{\,{\rm pc }}
\newcommand       \cm		{\,{\rm cm }}
\newcommand       \g		{\,{\rm g }}

\def \kms{\mbox{kms$^{-1}$}}

\def \cm2{\mbox{cm$^{-2}$}}
\def \cm3{\mbox{cm$^{-3}$}}

\title[Variations in the Galactic SFR]{Variations in the Galactic star formation
  rate and density thresholds for star formation}

\author[S.N.Longmore et al.]{\parbox{\textwidth}{S.~N.~Longmore$^{1}$\thanks{E-mail: \texttt{slongmor@eso.org}}, 
J.~Bally$^{2}$, 
L.~Testi$^{1,3}$, 
C.~R.~Purcell$^{4,5}$, 
A.~J.~Walsh$^{6}$,
E.~Bressert$^{1,7}$, 
M.~Pestalozzi$^{8}$, 
S.~Molinari$^{8}$, 
J.~Ott$^{9}$, 
L.~Cortese$^{1}$, 
C.~Battersby$^{2}$, 
N.~Murray$^{10}$, 
E.~Lee$^{10}$, 
D.~Kruijssen$^{11}$}\vspace{0.4cm}\\
$^{1}$European Southern Observatory, Karl-Schwarzschild-Strasse 2, D-85748 Garching bei M\"{u}nchen, Germany\\
$^{2}$Center for Astrophysics and Space Astronomy, University of Colorado, UCB 389, Boulder, CO 80309\\
$^{3}$INAF-Osservatorio Astrofisico di Arcetri, Largo E. Fermi 5, I-50125 Firenze, Italy\\
$^{4}$School of Physics and Astronomy, University of Leeds, Leeds LS2 9JT, UK\\
$^{5}$Sydney Institute for Astronomy (SiFA), School of Physics, The University of Sydney, NSW 2006, Australia\\
$^{6}$Centre for Astronomy, School of Engineering and Physical Sciences, James Cook University, Townsville, QLD, 4811, Australia\\
$^{7}$School of Physics, University of Exeter, Stocker Road, Exeter EX4 4QL\\
$^{8}$INAF-Istituto Fisica Spazio Interplanetario, Via Fosso del
   Cavaliere 100, I-00133 Roma, Italy\\
$^{9}$National Radio Astronomy Observatory, P.O. Box O, 1003 Lopezville Road, Socorro, NM 87801, USA\\
$^{10}$Canadian Institute for Theoretical Astrophysics, 60 St. George Street, University of Toronto,
Toronto, ON M5S 3H8, Canada\\
$^{11}$Max-Planck Institut fur Astrophysik, Karl-Schwarzschild-Strasse 1, 85748, Garching, Germany \\}
\begin{document}

\date{in prep.}
\pagerange{\pageref{firstpage}--\pageref{lastpage}} \pubyear{2012}
\maketitle


\begin{abstract}

  The conversion of gas into stars is a fundamental process in
  astrophysics and cosmology.  Stars are known to form from the
  gravitational collapse of dense clumps in interstellar molecular
  clouds, and it has been proposed that the resulting star formation
  rate is proportional to either the amount of mass above a threshold
  gas surface density, or the gas volume density.  These
  star-formation prescriptions appear to hold in nearby molecular
  clouds in our Milky Way Galaxy's disk as well as in distant galaxies
  where the star formation rates are often much larger.  The inner
  500\,pc of our Galaxy, the Central Molecular Zone (CMZ), contains
  the largest concentration of dense, high-surface density molecular
  gas in the Milky Way, providing an environment where the validity of
  star-formation prescriptions can be tested.  Here we show that by
  several measures, the current star formation rate in the CMZ is an
  order-of-magnitude lower than the rates predicted by the currently
  accepted prescriptions.  In particular, the region $1^\circ < l <
  3.5^\circ$, $|b| < 0.5^\circ$ contains $\sim10^7$\,M$_\odot$ of
  dense molecular gas --- enough to form 1000 Orion-like clusters ---
  but the present-day star formation rate within this gas is only
  equivalent to that in Orion. In addition to density, another
  property of molecular clouds, such as the amplitude of turbulent
  motions, must be included in the star-formation prescription to
  predict the star formation rate in a given mass of molecular gas.

\end{abstract}


\section{Introduction}
Stars play a pivotal role in shaping the cosmos. High-mass stars
contributed to the ionisation of the Universe after the Cosmic Dark
Ages.  They drive energy cycles and chemical enrichment on galactic
scales, hence sculpting galactic structure. Thus, the conversion of
gas into stars is fundamental to astrophysics and cosmology.

The rate at which gas is converted into stars has been measured in the
disks of nearby galaxies.  When averaged over hundreds of parsecs, the
star-formation rate (SFR) was found to have a power-law dependence on
the gas surface-density as described by the Schmidt-Kennicutt (SK)
relations \citep{schmidt1959,kennicutt1998,kennicutt_evans2012}.  A
linear relationship is found between SFR and gas surface-density above
a local extinction threshold, A$_{\rm K} \sim$0.8\, magnitudes at a
near-infrared wavelength of 2.2 $\mu$m, corresponding to a gas
column-density of $\sim7.4\times10^{21}$ hydrogen molecules per
cm$^{-2}$ or a gas surface-density $\Sigma_{\rm gas}$, of
$\sim116$\,M$_\odot$pc$^{-2}$ \citep{lada2012} .  For a typical 0.1 pc
radius cloud core, this corresponds to a volume density $ 3 \times
10^4$\, hydrogen molecules cm$^{-3}$.

Alternatively, it has been proposed that the SFR density is linearly
proportional to the mean gas density divided by the free-fall time
multiplied by an efficiency factor estimated to be about 1\%
\citep{kdm2012}.  These surface- and volume-density relations
potentially unify our understanding of SFRs from the nearest star
forming regions to ultra-luminous infrared galaxies (ULIRGs) and even
star forming regions at intermediate to high redshifts
\citep{swinbank2010,danielson2011}.

With the aim of testing these star formation (SF) scaling relations
across large volumes of the Milky Way (MW), in
$\S$~\ref{sec:dg_sf_tracers} we start by looking at the choice of
observational diagnostics of dense gas and star formation activity in
the Galaxy. In $\S$~\ref{sec:obs_data}, we present the survey data
used in this work. In
$\S$~\ref{sec:discussion_sfr_per_unit_mass_in_gal}~
\&~\ref{sec:variation?} we discuss the striking differences between
the dense gas and SF activity tracers across the Galaxy and
investigate if any observational or systematic biases may be causing
such differences. We find no evidence that the results are strongly
affected by observational or systematic biases. In
$\S$~\ref{sec:quant_analyis_cmz} we then perform a quantitative
analysis of the SFR and gas mass in the CMZ and directly compare this
to predictions of proposed column density threshold and volumetric SF
relations. In $\S$~\ref{sec:summary} we summarise the results and
discuss the implications for the universality of star formation
relations.

\section{Observational diagnostics of dense gas and star formation activity in the Milky Way}
\label{sec:dg_sf_tracers}

We aim to test star formation scaling relations for representative
samples of SF regions across large volumes of the MW. Such systematic
studies are now possible thanks to the (near) completion of many
blind, multi-wavelength, continuum and spectral line Galactic plane
surveys. However, care must be taken in the choice of SF activity
diagnostics. All surveys aiming to observe a large fraction of the
dense gas in the Galaxy must look directly through the Galactic disk
plane. Extinction therefore makes traditional UV, optical and even
near-IR diagnostics of SF activity unusable for all but the closest
regions. 24$\mu$m emission, commonly used to trace SF in extragalactic
studies, suffers from extinction only in extreme environments (like
the GC) but contamination from evolved stars can be significant,
requiring other wavelength data to determine the source
nature. Sensitivity and completeness limits mean counting young
stellar objects still heavily embedded in their natal molecular clouds
can only be done reliably for the closest SF regions.

Diagnostics are required at wavelengths longer than a few hundred
microns, for which the Galaxy is effectively optically-thin. Sub-mm
dust continuum emission is a robust gas mass tracer, but from this
alone it is difficult to impose a volume density cutoff on the sample
selection. Molecular line emission has the advantage that it can be
used to probe gas within a certain density range. The actual density
of gas probed by observations of a given transition is related to the
critical density of that transition (the density at which the
collisional de-excitation rate equals the Einstein A coefficient), but
also depends non trivially on the sensitivity of the observations, the
line opacity and gas kinetic temperature. A useful concept, taking
into account these affects, is the effective critical density
\citep{evans1999}.  For a given transition, this is the density of gas
probed by observations of a particular brightness temperature
sensitivity for gas at a given kinetic temperature. Assuming detected
emission comes from gas at a similar kinetic temperature (a reasonable
assumption for dense gas clumps in molecular clouds) and that beam
dilution is not an issue, a uniform sensitivity molecular line
transition survey offers a natural way to probe the amount of gas at,
and above, the effective critical density.

Masers are well known indicators of SF activity and with strong
transitions in the cm/mm wavelength regime they can be detected at
large distances. However, it is not currently possible to derive an
absolute star formation rate from first principles using
masers. Free-free emission from ionised gas in \hii\ regions created
by young high mass stars is readily traced by cm-continuum and radio
recombination line (RRL) emission, both of which are readily
observable over large wavelength/frequency ranges
\citep{peters2012}. The number count of \hii\ regions can give a
handle on the relative numbers of SF sites across the Galaxy. The
absolute star formation rate can be estimated from the luminosity of
the cm-continuum emission by calculating the rate of ionising photons
from high-mass stars \citep{murray_rahman2010}, although contamination
from non-thermal emission (which can lead to an overestimate of the SF
activity) must be taken into account, especially towards the GC.  It
should be noted that each of these SF diagnostics traces the SF
activity over different timescales. Masers are only observed towards
regions that are still actively forming stars, so signpost SF activity
on timescales of a few 10$^5$yr. Free-free emission from \hii\ regions
is observed over the lifetimes of the high mass stars providing the
ionising photons ie. timescales of a few Myr
\citep{murray_rahman2010}. When we refer to the SFR throughout this
work, we are implicitly talking about star formation activity over
these timescales, which is significantly shorter than other
diagnostics (e.g. UV emission).

Of the recent and planned surveys\footnote{Although other large
  Galactic, molecular-line mapping projects are also being conducted
  on the Mopra telescope
  e.g. \citet{barnes2011,foster2011,jones2012}}, the H$_2$O Galactic
Plane Survey \citep[HOPS:][]{walsh2008,walsh2011,purcell2012} is the
only blind molecular line survey covering a large fraction (100
deg$^2$) of the Galactic plane simultaneously in multiple thermal
molecular lines (including the important $\nhthree$ molecule), masers
and RRLs . Below we describe HOPS and several newly available Galactic
plane surveys (see Table~\ref{tab:surveys}) which we use to
investigate how the SFR relations and proposed density thresholds hold
for more representative SF regions across the MW.

\section{Observational data}
\label{sec:obs_data}

Determining the rate of SF per unit mass of gas requires measuring, i)
the total gas mass and, ii) how much SF is taking place within this
gas. We make use of recently available $\nhone$ data from HOPS and 70
to 500\,$\mu$m data from the Herschel Infrared Galactic Plane Survey
\citep[HIGAL:][]{molinari2010} to trace the mass of dense gas across
the Galaxy. Ammonia ($\nhthree$) is well known as an excellent tracer
of dense molecular gas at all evolutionary phases of the SF process
\citep{ho_townes1983,longmore2007,beuther2009,hill2010,longmore2011},
including the earliest phases where low temperatures mean molecules
like CO freeze out on to dust grains and no longer reliably trace the
gas. Given the HOPS sensitivity of 0.2\,K we expect to detect gas with
an effective critical density of a few 10$^3$, probing gas close to,
and above, the proposed volume density threshold
\citep{evans1999}. The 500\,$\mu$m HiGAL data traces optically-thin
dust emission, providing a measurement of the gas column density along
a given line of sight. To directly compare the 500$\mu$m and $\nhone$
emission, we first re-binned the 35$\arcsec$ 500$\mu$m map to the HOPS
pixel scale using Montage (http://montage.ipac.caltech.edu/), then
applied the HOPS $\nhone$ emission mask to the re-binned HiGAL
map. This provided the 500$\mu$m emission from all HOPS $\nhthree$
emission regions.

We use data from HOPS and the methanol multibeam
\citep[MMB][]{caswell2010,caswell2011,green2010,green2012} surveys to
detect the 22\,GHz water maser emission and 6.7\,GHz Class II methanol
maser emission, respectively, across the Galactic plane. Both water
and methanol masers are well known indicators of SF activity within
molecular clouds. In particular, the 6.7\,GHz Class II methanol maser
transition is known to exclusively trace regions of high mass
($>$8\,M$_\odot$) \citep{minier2003}\footnote{Although the origin of
  the masers in disks or outflows is debated
  \citep[e.g.][]{norris1998,minier2000,debuizer2009,debuizer2012}.}. Water
masers are detected towards evolved stars, but these are found outside
the disk-plane and the emission is typically two orders of magnitude
weaker \citep{caswell2010}. So sensitivity-limited, water maser
Galactic plane surveys should predominantly trace SF regions. Removing
evolved star contaminants is typically straightforward using other
readily-available multi-wavelength data \citep{lumsden2002}.

We use the radio recombination line (RRL) data from HOPS and the Green
Bank Telescope (GBT) \hii\ Region Discovery Survey
\citep[HRDS:][]{anderson2011} to compare the Galactic \hii\ region
number distribution to the dense gas and maser distributions. The HRDS
completeness limit is sufficient to detect all \hii\ regions ionised
by single O-stars to a distance of 12 kpc. Both HOPS and HRDS use RRLs
to detect emission from \hii\ regions and the kinematic dimension
helps to mitigate line-of-sight confusion that can affect continuum
surveys. HRDS is much more sensitive than HOPS but is a targeted
survey based on previous large-area cm-continuum and IR surveys. The
HOPS RRL emission is less sensitive but allows for direct comparison
with the maser, $\nhthree$ and 500\,$\mu$m data over the full
100\,deg$^2$ of the HOPS survey region.

The \hii\ region number counts provide a measure of the relative
distribution of SF sites across the Galaxy but we make use of recent
analysis of WMAP data \citep{murray_rahman2010} to derive the absolute
SFR in the GC.

\section{Comparing dense gas emission and SF activity tracers across the Galaxy}
\label{sec:discussion_sfr_per_unit_mass_in_gal}

As the Galaxy is optically-thin to the $\nhone$, 500\,$\mu$m, water
maser, methanol maser and RRL emission, the emission surface density
ratios between these tracers in the plane of the sky is equivalent to
that in a face-on view of the Galaxy. Comparing the emission ratios as
a function of longitude and making the reasonable assumptions that i)
the integrated intensity of the far-IR and $\nhone$ emission is
proportional to the mass of dense gas, and, ii) the number of water
masers, methanol masers, \hii\ regions and RRL integrated intensity is
proportional to the amount of SF, we now investigate the relationship
between the dense gas and SFR in the Galaxy.

\begin{figure*}
\begin{center}
\includegraphics[width=0.9\textwidth, angle=0, trim=0 0 -5 0]{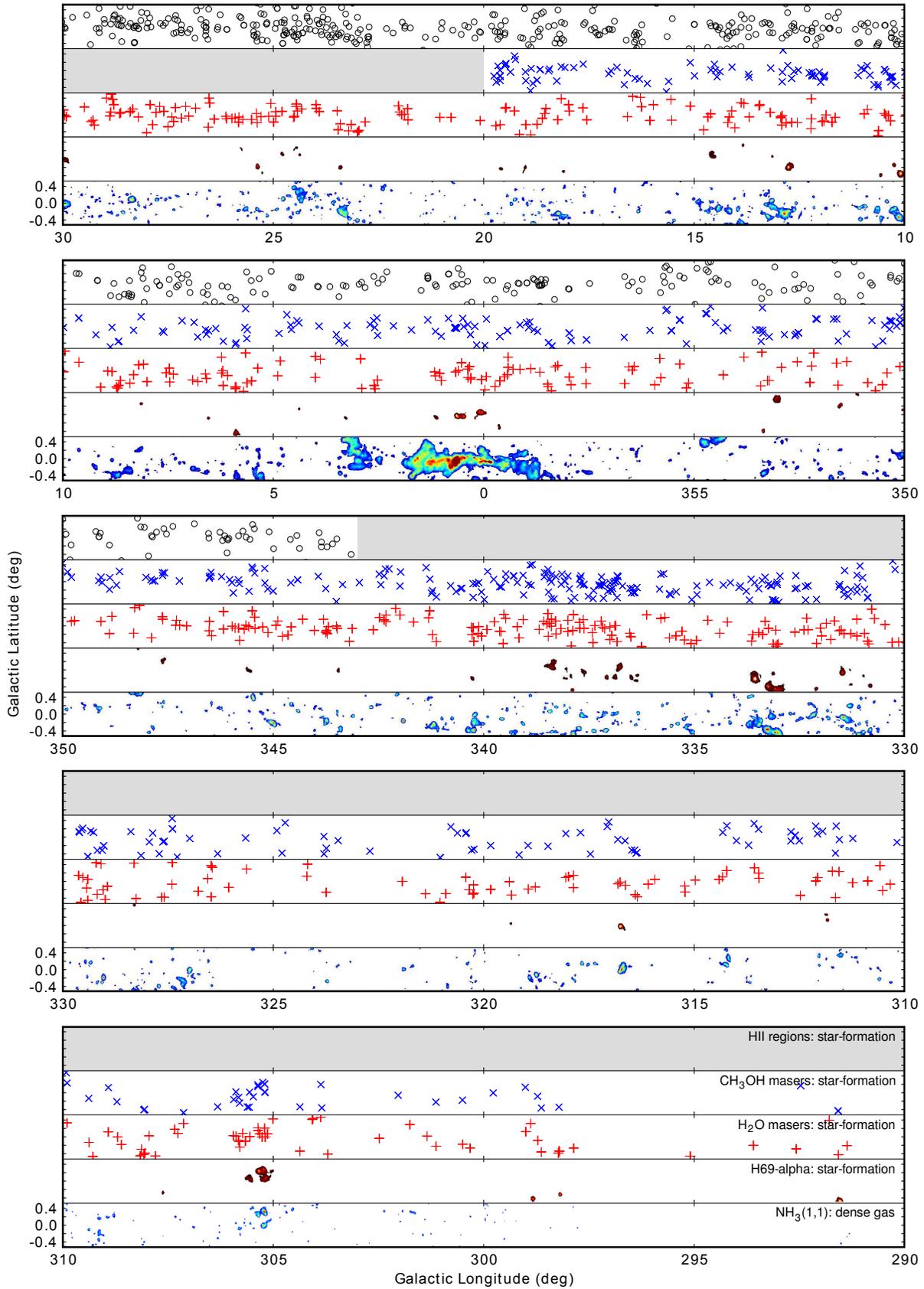} \\
\end{center}
\caption{\small{Distribution of dense gas and star formation activity
    tracers as a function of Galactic latitude and longitude. Black
    circles mark the positions of \hii\ regions, blue crosses show
    methanol masers and red plus symbols mark water masers. Regions of
    sky not covered in these surveys are shaded in grey. $\nhone$
    [bottom] and H69$\alpha$ [second-bottom] integrated intensity
    emission is displayed using a square-root image stretch. The
    tracers and their function as either a dense gas or star formation
    activity tracer are labelled at the right hand edge of the bottom
    row. The central molecular zone (CMZ) can be seen as bright,
    extended $\nhone$ emission from longitudes of roughly 358$^\circ$
    to 4$^\circ$.}}
\label{fig:long_nh3_mas}
\end{figure*}

Figure~\ref{fig:long_nh3_mas} shows the dense gas emission and current
star formation activity tracers as a function of Galactic longitude
($l$) and latitude ($b$).  The dense gas distribution is dominated by
the very bright and spatially-extended emission within a few degrees
of the Galactic centre -- the CMZ \citep{morris_serabyn1996,
  ferriere2007}. This is easily distinguished from the rest of the
Galactic plane by the very high intensity of dense gas emission and
large gas velocity dispersion as indicated by wide spectral
lines. While the dense gas emission is highly concentrated in the CMZ,
the distribution of star formation activity tracers is relatively
uniform across the Galaxy. Quantitatively, the CMZ accounts for
$\sim$80\% of the integrated $\nhone$ intensity but only contains 4\%
of the survey area.  Yet, the CMZ does not stand out in either water
or methanol maser emission, or in the number of \hii~regions, which
all trace recent high-mass star formation. A qualitatively similar
trend is reported by \citet{beuther2012} who compare the sub-mm dust
continuum emission and GLIMPSE point sources as a function of Galactic
longitude.

\begin{figure*}
\begin{center}
\includegraphics[width=0.8\textwidth, angle=0]{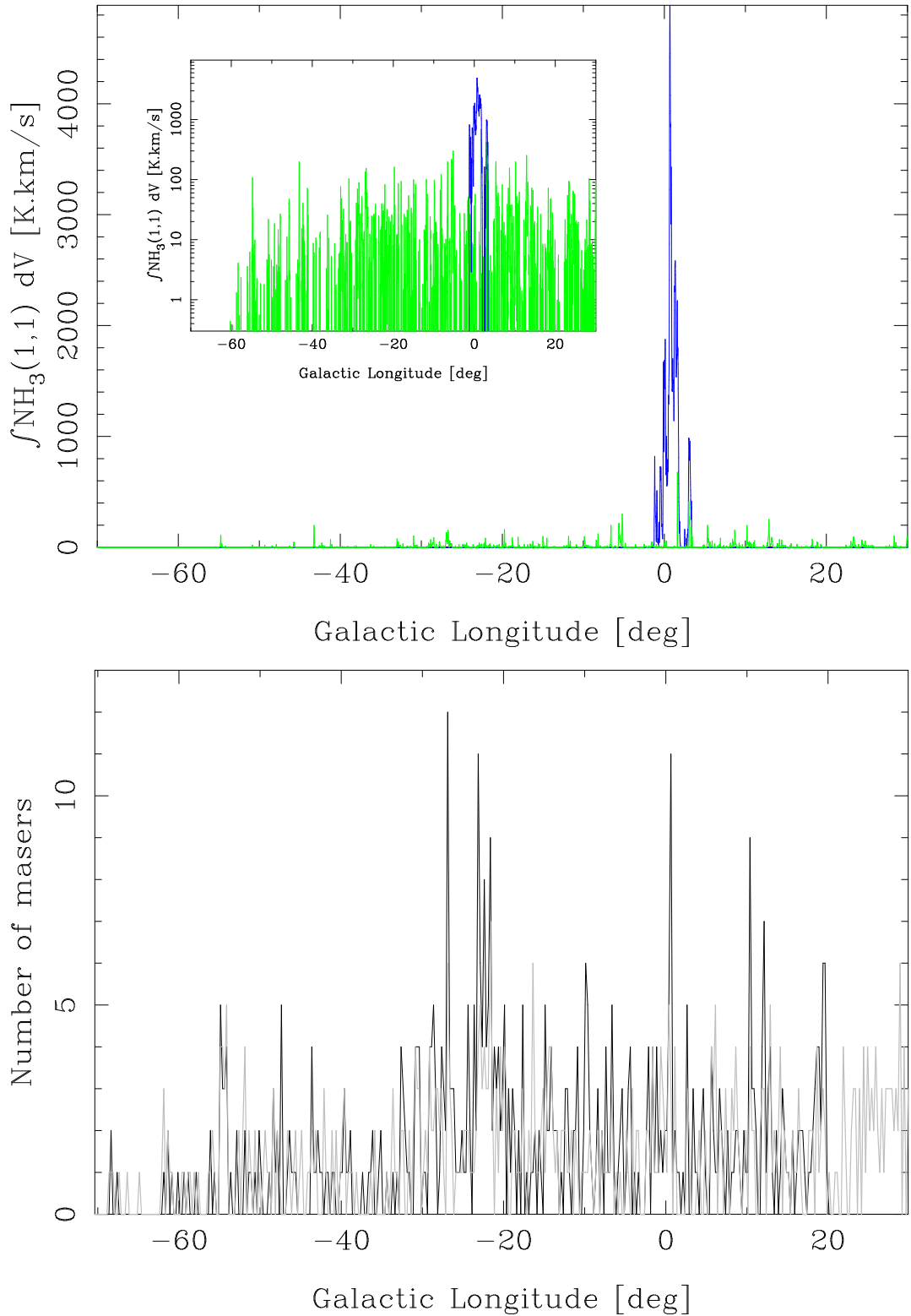} \\
\includegraphics[height=0.8\textwidth, angle=-90]{fig2b.ps} \\
\end{center}
\caption{\small{[Top] Dense gas distribution, traced by the $\nhone$
    integrated intensity, as a function of Galactic longitude
    (2$^\prime$ bins). The emission was split into two regions: a
    Galactic centre region encompassing the central molecular zone
    (blue) and the rest of the Galaxy (green). The Galactic centre
    region clearly dominates, accounting for $\sim$80\% of the
    $\nhone$ integrated intensity emission, despite only accounting
    for $\sim$4\% of the total survey area. The inset shows the same
    plot but with a logarithmic scale on the y-axis. [Bottom] Star
    formation activity, traced by the number of water masers (grey)
    and methanol masers (black: data for $l>$20$^\circ$ unavailable),
    as a function of Galactic longitude (0.25$^\circ$ bins). In
    comparison to the $\nhone$ emission, the distribution of both
    maser species is relatively flat with Galactic longitude. }}
\label{fig:long_nh3_mas_hist}
\end{figure*}

Figure~\ref{fig:long_nh3_mas_hist} shows Galactic longitude
distributions of dense gas and star-formation activity indicators
summed over the observed latitude range ($-0.5^\circ \leq b \leq
0.5^\circ$) for each longitude pixel. To make a direct comparison of
dense gas and SF tracers as a function of longitude, we first sampled
in 2-degree longitude bins to ensure the volume of the Galaxy covered
in each bin contains a large number of SF regions and so is
appropriate for testing SF relations \citep{onodera2010}. The emission
or number count in each 2 degree longitude bin was then normalised by
the total emission or number in the full survey area.

The top panel of Figure~\ref{fig:dense_gas_vs_masers} shows that
there is a strong correlation between independent dense gas
tracers. To separate the CMZ emission from that in the rest of the
Milky Way, from here on we refer to the region $|l|<5^\circ$ with
$\nhone$ line widths $\Delta$V$~>$~15\,$\kms$ as ``GC-only'', to
distinguish it from the rest of the Galaxy, which we refer to as the
``non-GC'' region. No offset is seen in the correlation between the
GC-only and non-GC regions. The middle panel of
Figure~\ref{fig:dense_gas_vs_masers} shows a correlation between the
dense gas and SF tracers for the non-GC regions. However, the GC-only
regions are clearly distinct, with at least an order of magnitude
brighter dense gas emission for the number of SF activity tracers.

The bottom panel of Figure~\ref{fig:dense_gas_vs_masers} shows the
resulting dense gas vs SF tracer surface density ratio in
Galacto-centric radius, R$_{\rm GC}$, annuli of 0.5\,kpc. R$_{\rm GC}$
was calculated using the Galactic rotation curve of
\citet{brand_blitz1993} and assuming a distance to the Galactic centre
of 8.5 kpc and a solar velocity of 220\,kms$^{-1}$. The gas between
the CMZ and R$_{\rm GC} \sim$3\,kpc shows emission at anomalous
velocities so the rotation curve does not place reliable constraints
on R$_{\rm GC}$ over this range. The surface density ratios over this
region, highlighted by the hatched rectangle, should be ignored. The
ratio is approximately constant at R$_{\rm GC}$$>$3\,kpc. This
suggests the linear relationship observed between the quantity of gas
above the proposed extinction threshold and the SFR
\citep{gao_solomon2004,wu2005,lada2012} extends to a larger number of
more representative SF regions across the Galaxy. By comparison, the
dense gas surface density towards the GC (R$_{\rm GC}$$<$0.5\,kpc) is
orders of magnitude too large compared to the SF activity surface
density.

\section{Variation in SFR per unit mass of dense gas between the disk and Galactic centre?}
\label{sec:variation?}

We now investigate the striking difference between the longitude and
surface density distribution of dense gas and SF tracers between the
GC-only and non-GC regions (Figure~\ref{fig:dense_gas_vs_masers}). If
the number of masers and \hii\ regions were directly proportional to
the integrated intensity of the dense gas emission, as would be
expected if assumptions i) and ii) outlined in
$\S$~\ref{sec:discussion_sfr_per_unit_mass_in_gal} hold, and the SFR
was set by the amount of dense gas, how many would be expected towards
the GC-only region? With $\sim$1/25 of the HOPS survey area and 4
times the total $\nhone$ integrated intensity outside the GC, the
GC-only region has $\sim$100 times the dense gas integrated intensity
per unit survey area compared to the non-GC region. Therefore, the
expected number of masers and \hii\ regions towards the GC should also
be 100 times larger per unit area. The SF activity per unit area as
traced by methanol masers is a factor 3 to 6 higher in the inner
250\,pc of the Galaxy compared to the spiral arms \citep{caswell2010}.
However, our results show that this increase in maser surface density
is dwarfed by the increase in the dense gas surface density. Given the
amount of dense gas, the GC-only region appears deficient in SF
tracers by two orders of magnitude. Taken at face value this suggests
large differences in SFR per unit mass of dense gas between the disk
and Galactic centre. Below we investigate if such a large deficit
could be caused by observational or systematic biases affecting the
emission from one or more of the tracers used in this study.

\subsection{Potential observational biases}
\label{sub:result_obs_bias}

These data are from blind large-area surveys with approximately
uniform sensitivity. Therefore, the results should not be affected by
target selection criteria or variable sensitivity. But what about the
sensitivity limits themselves?  For example, could HOPS be missing a
large fraction of the dense molecular gas outside the GC-only region
as this $\nhone$ emission falls beneath the sensitivity threshold?
\citet{purcell2012} show HOPS should detect 400\,M$_\odot$ clumps to
5.1 kpc and 5000\,M$_\odot$ clumps to the Galactic centre
distance. These limits suggest completeness is not a problem.

Alternatively, as the GC-only clouds are more distant on average than
the non-GC regions, could there be a large number of weak water masers
towards the GC-only region which fall below the HOPS detection limit?
Ott et al (in prep.) have mapped the entire CMZ from 20 to 28\,GHz
with Mopra in a similar setup to the HOPS observations but with
$\sim8\times$ better sensitivity and using the correlator in broadband
rather than zoom mode (resulting in velocity resolution of 3.5\,$\kms$
rather than 0.4\,$\kms$ as for HOPS). They find approximately twice
the number of masers -- not enough to account for the apparent
deficit.

\begin{figure*}
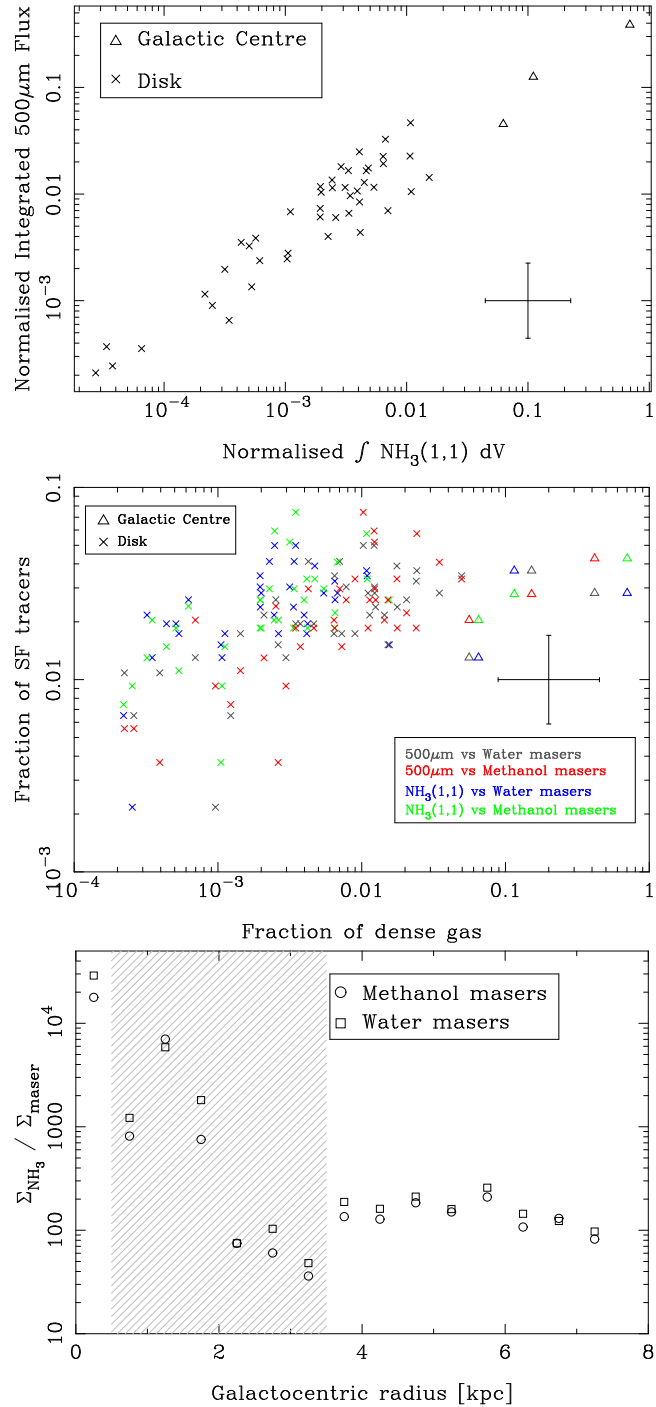

\begin{center}
\includegraphics[height=8.5cm, angle=-90, trim=0 0 -5 0]{fig3a.ps} \\
\includegraphics[height=8.5cm, angle=-90, trim=0 0 -5 0]{fig3b.ps} \\
\includegraphics[height=8.5cm, angle=-90, trim=0 0 -5 0]{fig3c.ps} \\
\end{center}
\caption{\small{[Top] Comparison of two independent dense gas tracers,
    $\nhone$ integrated intensity vs. the integrated 500\,$\mu$m flux,
    for each 2$^\circ$ longitude bin (normalised to the total flux in
    that tracer). [Middle] The fraction of dense gas tracers ($\nhone$
    integrated intensity and 500$\mu$m flux) vs fraction of star
    formation tracers (water and methanol masers) for each 2$^\circ$
    longitude bin between $-70^\circ<l<20^\circ$ (the
    mutually-overlapping coverage of MMB and HOPS). Different colours
    show different combinations of dense gas and SF tracers. In both
    top and middle plots, the Galactic centre (GC-only) and disk
    (non-GC) regions are shown as triangles and crosses,
    respectively. Representative uncertainties of a factor 5 in dense
    gas mass and 3 in maser counts are shown as error bars. [Bottom]
    Ratio of dense gas surface density ($\nhone$ integrated intensity)
    to star formation surface density (methanol and water masers) as a
    function of Galacto-centric radius, R$_{\rm GC}$. The hatched
    region shows the radial range over which the rotation curve used
    to derive R$_{\rm GC}$ is unreliable. The surface density ratios
    over this region should be ignored.}}
\label{fig:dense_gas_vs_masers}
\end{figure*}

Sites of maser emission observed with a single dish may be resolved
into multiple sites when observed at higher angular resolution with an
interferometer. This will not affect the methanol masers counts as
they all have interferometric follow-up, but may affect the HOPS water
maser distribution. To investigate this, we compared the HOPS
detections to $\sim$4$\times$ deeper, interferometric water maser
observations toward a 0.5\,deg$^2$ region surrounding the GC
\citep{caswell2011b}. In the same region ($|l|,|b|<0.4^\circ$), 27
masers were found compared to the 8 found by HOPS, and 3 of the 8 HOPS
detections were resolved into 2 maser sites. If this is representative
of the whole GC-only region HOPS may have underestimated the number of
water masers by a factor of 3. Preliminary results from a deeper,
interferometric EVLA water maser survey covering $|l|<1.5^\circ$,
$|b|<1^\circ$ also recover a similar factor of 3 increase in number of
detections compared to HOPS over the same region (J. Ott,
F. Yusef-Zadeh private communication). This is not sufficient to
explain the large disparity between the maser and dense gas
distributions in the GC-only and non-GC regions. Just as importantly,
there is also no reason to expect a systematic difference between the
number of maser sites each HOPS detection will be resolved into
between the GC and non-GC regions. This is confirmed by preliminary
results from the interferometric follow-up observations of the HOPS
masers (Walsh et al, in prep). As a further check, repeating the
analysis using the integrated intensity of the masers (which will not
be affected by how many maser sites each HOPS detection is resolved
into at higher resolution), rather than number of masers, as a
function of longitude produces the same results in
Figures~\ref{fig:long_nh3_mas_hist}~\&~\ref{fig:dense_gas_vs_masers} .
We adopt a conservative factor of 3 uncertainty in the maser counts,
which we illustrate by the error bar in the middle panel of
Figure~\ref{fig:dense_gas_vs_masers}. The similarity of both maser
distributions in the bottom panel of
Figure~\ref{fig:dense_gas_vs_masers} gives confidence in the
robustness of the maser counts and the adopted uncertainty.

Finally, the trends in Figure~\ref{fig:dense_gas_vs_masers} consist of
ratios between observed parameters rather than the absolute values of
the parameters themselves. The distance-dependence on expected source
flux is the same for all observed parameters, so the trend itself will
not be affected by systematic differences in distance between the
GC-only and non-GC regions. We conclude that there is no evidence that
our results may be strongly affected by observational biases.

\subsection{Potential systematic biases in the dense gas and SF tracers}

Many mechanisms can affect assumption i) above, the most important of
these being: the emission optical depth (assumed to be
optically-thin); ammonia abundance and gas excitation conditions; dust
properties (composition, temperature) and gas-to-dust ratio. These
properties will certainly vary from one molecular cloud to another and
undoubtedly cause much of the scatter seen in
Figure~\ref{fig:dense_gas_vs_masers} as well as the non-linear slope
in the $\nhone$ integrated intensity vs 500$\mu$m flux. The variation
in these properties must be understood before interpreting differences
between individual clouds or deriving physical properties of the gas
directly from one of these tracers. However, we are averaging over
many molecular cloud complexes and are interested in
order-of-magnitude variations across the Galaxy so focus on systematic
differences between GC-only and non-GC regions.

\subsubsection{Potential systematic biases in the dense gas mass estimates}

Regarding the mass estimates, the emission optical depth will on
average likely be much higher towards the GC-only regions, leading to
a substantial underestimate of the dense gas mass towards the GC. This
only accentuates the difference between the Galactic centre and the
rest of the Galaxy. However, many observations show the GC excitation
conditions and dust temperature are on average higher. The most
relevant temperature measurements for this work are those derived
directly from $\nhthree$. Targeted observations of multiple $\nhthree$
inversion transitions towards GC GMCs show these have at least two
kinetic temperature regimes -- a cool (25\,K) component which
dominates the column density and a warm (200\,K) component which
accounts for $\sim$25\% of the column density
\citep{huttemeister1993}. Beam-averaged temperature measurements at
the HOPS resolution results in an average $T_K$ of $\sim$50\,K for GC
clouds (J. Ott, priv comm.). Compared to $\sim$20\,K for typical GMCs,
this would lead to an overestimate of the GC dense gas mass by a
factor $2-3$.

As the average gas kinetic temperature is higher in the GC, the
$\nhone$ effective critical density will be lower. Given the uniform
sensitivity of the HOPS observations, this means the GC $\nhone$
integrated intensity will include emission from lower density gas. As
the ultimate goal is to compare the amount of gas across the Galaxy
above a single density threshold, including lower density material
will lead to a systematic overestimate of this value towards the
GC. The magnitude of the overestimate will depend on both the
difference between the effective critical density probed towards the
GC and the rest of the Galaxy, and how much additional gas in the GC
lies between these two effective critical density limits. The $\nhone$
effective critical density changes by less than a factor 2 for gas
with kinetic temperatures between 10\,K and 100\,K
\citep{evans1999}. The line brightness temperature of 1\,K used in
these calculations corresponds to 5 times the HOPS r.m.s. sensitivity
so is a sensible limit. As the average gas kinetic temperature
difference between the GC and the Galaxy is much less than this (see
above), the effective critical density probed will be very similar. In
$\S$~\ref{sec:quant_analyis_cmz} we estimate the average density of
gas in the CMZ to be $\sim5\times10^3$\,cm$^{-3}$, so most of the gas
will be close to or above the effective critical density. We conclude
systematic differences in the effective critical density probed
between the GC-only and non-GC regions will not affect the result.

Extrapolating the observed metallicity gradient in the disk of $-0.03$
to $-0.07$\,dex\,kpc$^{-1}$ \citep{balser2011}, one would expect the
metallicity to increase by a factor 3$-$4 from the sun to the Galactic
centre. While some studies do find an increased metallicity towards
the Galactic centre, others find close to solar values
\citep{shields_ferland1994,najarro2009}. Using the commonly adopted
assumption that the Galactic centre metallicity is twice solar
\citep{ferriere2007}, the gas-to-dust ratio in the Galactic centre is
likely lower by the same value. This means the 500$\mu$m flux towards
the Galactic centre systematically over-estimates the dense gas mass
by a factor $\sim$2 compared to the disk. Similarly, the $\nhthree$
abundance gradient in the disk \citep{dunham2011} implies an average
factor $\sim$2 relative overestimate in the Galactic centre
$\nhthree$-derived mass estimate. Both the 500$\mu$m and $\nhthree$
dense gas mass estimates therefore over-predict the Galactic centre
mass by a factor $\sim$2. No systematic effects are known regarding
dust composition with Galacto-centric radius.

\begin{figure*}
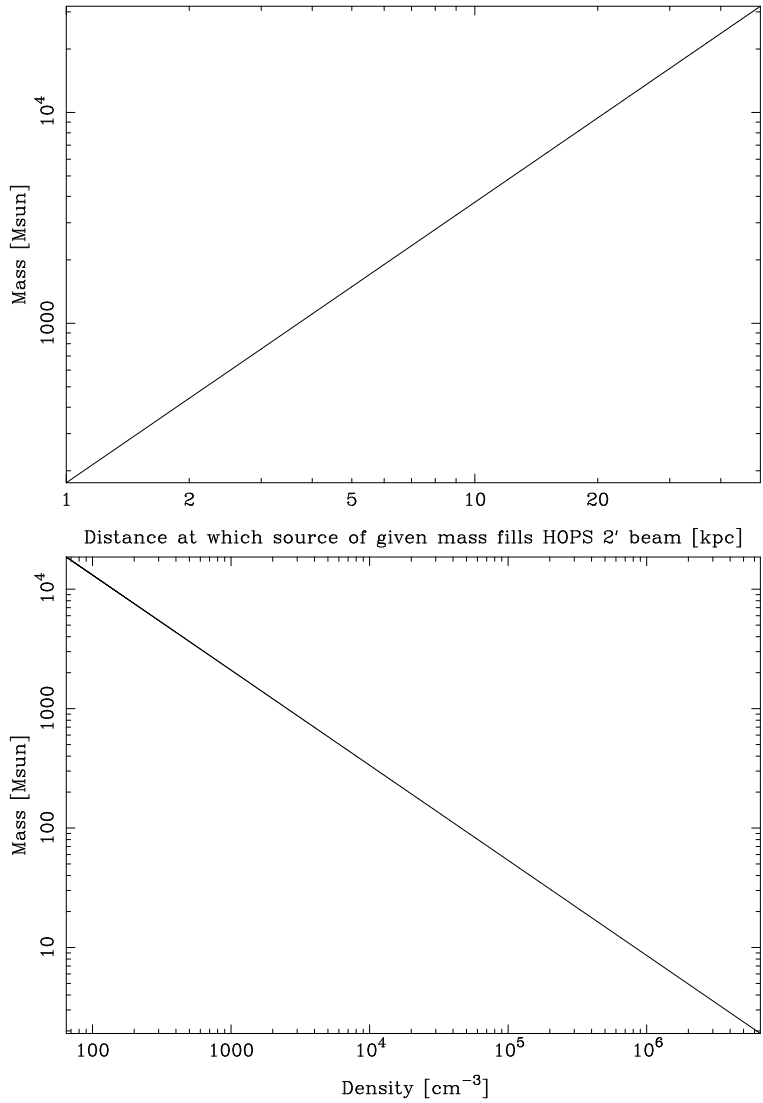

\begin{center}
\includegraphics[height=10cm, angle=-90, trim=0 0 -5 0]{fig6a.ps} \\
\includegraphics[height=10cm, angle=-90, trim=0 0 -5 0]{fig6b.ps} \\
\end{center}
\caption{\small{[Top] Estimating the distance at which molecular
    clouds of a given mass will be the same angular size as the
    2$^\prime$ HOPS beam. The physical radius as a function of mass
    was estimated using the \citet{kauffmann_pillai2010} empirical
    mass-size relationship for molecular clouds which will proceed to
    form high mass stars. The line in the plot shows the distance at
    which a molecular cloud of that mass will have an angular size of
    2$^\prime$. Molecular clouds below the line will suffer from beam
    dilution so the observed surface brightness, and therefore
    integrated intensity, will be reduced. Molecular clouds above this
    line will not suffer from this affect. [Bottom] Density of clouds
    as a function of their mass, calculated from the
    \citet{kauffmann_pillai2010} empirical mass-size relationship. }}
\label{fig:mass_dist_kp_hops_resolve}
\end{figure*}

Another potential worry is related to beam dilution. The $\nhone$
emission from the GC covers a large angular area on the sky and thus
fills the 2$^\prime$ beam of the HOPS observations. If the molecular
clouds outside of the GC systematically had angular sizes much smaller
than the 2$^\prime$ beam, the measured $\nhone$ surface brightness
would be affected by beam dilution and the integrated intensity would
be systematically underestimated. To assess whether this affect is
important, we estimated the distance at which molecular clouds of a
given mass will be the same angular size as the 2$^\prime$ HOPS
beam. The physical radius as a function of mass was estimated using
the \citet{kauffmann_pillai2010} empirical mass-size relationship, $M
= 870 M_\odot(r/pc)^{1.33}$, for molecular clouds which will proceed
to form high mass stars. Based on this relation, the line in the top
panel of Figure~\ref{fig:mass_dist_kp_hops_resolve} shows the distance
at which a molecular cloud of a given mass will have an angular size
of 2$^\prime$. Molecular clouds below the line will suffer from beam
dilution so the observed surface brightness will be reduced, and the
integrated intensity underestimated. From this we conclude that beam
dilution is undoubtedly an important factor for low-mass clouds at
large distances. An implicit assumption in the above analysis is that
it is possible to detect the emission from all the gas in the
cloud. However, the density fluctuations in molecular clouds mean that
some of the gas may lie at low density and not excite $\nhthree$
emission. We can estimate the effect of this by calculating the
average density of the gas inferred from the empirical mass-size
relation \citep{kauffmann_pillai2010}, $ n(H_2) = 3.13\times10^3
(r/pc)^{-1.67}$ cm$^{-3}$. As shown in the bottom panel of
Figure~\ref{fig:mass_dist_kp_hops_resolve}, clouds up to
$\sim$1000\,M$_\odot$ have average volume densities several
10$^3$cm$^{-3}$ and above, so we would expected most of the gas to
emit in $\nhthree$. The average density of more massive clouds drops,
so the fraction of $\nhthree$ emitting gas will drop too. Although the
top panel of Figure~\ref{fig:mass_dist_kp_hops_resolve} shows HOPS
will easily resolve very massive clouds to large distances, the
$\nhthree$ emission must come from denser clumps within these clouds.
The extent to which beam dilution affects the detected $\nhthree$
emission, if at all, depends on the density fluctuations within the
clouds, which will vary from cloud to cloud. However, the fact that in
general the larger the cloud, the better it will be resolved,
mitigates the fact that less of the gas will be at high density.  We
conclude that while beam dilution undoubtedly affects low-mass clouds
at large distances, it is not an important issue for clouds of a
thousand solar masses and greater. It is the clouds in this mass
regime that will be forming the high-mass stars traced by methanol
maser emission and HII regions.

On balance, the systematic effects would argue that making assumption
i) would overestimate the dense gas mass towards the GC relative to
the rest of the Galaxy. However, it seems highly unlikely these would
lead to an overestimate by a factor of $\sim$100. Additionally, as
most of these effects are (at least partially) independent between the
500\,$\mu$m and ammonia emission, the fact that no jump is seen in the
correlation between the two tracers in
Figure~\ref{fig:dense_gas_vs_masers} when comparing the GC-only and
non-GC regions argues the systematics are not important for the
analysis presented here. Based on the above arguments and the observed
scatter in the 500\,$\mu$m vs ammonia emission correlation (top panel
of Figure~\ref{fig:dense_gas_vs_masers}), we adopt a factor 5
uncertainty, which is illustrated by the error bars in
Figure~\ref{fig:dense_gas_vs_masers}.

\subsubsection{Potential systematic biases in the maser counts}

As water masers are also observed towards evolved stars, contamination
may affect the maser number counts. As previously stated, the water
masers from evolved stars are found outside the disk-plane and the
emission is typically two orders of magnitude weaker
\citep{caswell2010}. Therefore, sensitivity-limited, water maser
Galactic plane surveys should predominantly trace SF
regions. Nevertheless, we are conducting high angular resolution
follow-up observations of the HOPS masers to uncover the true nature
of each maser detection. Preliminary results show the majority
($\sim$70\%) are associated with SF. Also, no trends in the relative
number of masers associated with evolved stars or SF regions are seen
with Galactic longitude (Walsh et al, in prep). In any case, if
anything, one might expect more evolved stars towards the bulge and
Galactic centre. Contamination of water masers from evolved stars
would lead to an \emph{increase} in number of water masers observed
towards the GC. This only strengthens the result. Finally, the
similarity of the water maser distribution in the bottom panel of
Figure~\ref{fig:dense_gas_vs_masers} to the methanol maser
distribution (which are not observed towards evolved stars), adds
further weight that contamination does not affect the results.

Given the tight constraints on the physical conditions required for
masers to exist, another possibility is that maser emission is somehow
quenched towards the Galactic centre region. This seems plausible
given the extreme environmental conditions in the Galactic centre
compared to the rest of the disk \citep[e.g. interstellar radiation
  field enhanced by 10$^3$, external pressure P/k$\sim$10$^8$, strong
  magnetic fields, turbulence, gamma
  rays:][]{morris_serabyn1996,ferriere2007,crocker2010}. However,
masers are formed deeply embedded within GMCs so they are shielded
from the interstellar radiation field. Regarding the effect of
turbulence, the coherence path lengths for masers are of order AU size
scales, so individual maser emission regions will not be affected by
velocity flows or gas kinematics on pc-scales (Mark Reid, private
communication). Additionally, extremely bright ``mega masers'' are
known to exist in the circumnuclear disks around supermassive black
holes -- an even more extreme environment than the CMZ. It is worth
pointing out that the pumping mechanisms for water and methanol masers
are different. As far as we are aware, no known effect in maser
pumping theory can account for the observed factor 100 deficit in the
number of both water and methanol masers towards the GC-only
region. We conclude there is no evidence that our results are strongly
affected by systematic biases in the dense gas and SF tracers.

\section{Quantitative analysis of the SFR and gas mass in the CMZ}
\label{sec:quant_analyis_cmz}

The previous qualitative analysis strongly points to large differences
between the SFR per unit mass of dense gas in the CMZ and the rest of
the Galaxy. However, the inability to derive absolute SFRs from first
principle for masers and the large uncertainties in deriving masses
from $\nhone$ integrated intensities preclude us from a more
quantitative analysis using these tracers. Therefore, we now focus on
the CMZ, the region which stands out as potentially discrepant, and
aim to directly test the predictions of different SF relations.

\subsection{Structure and physical properties of the molecular gas
  in the Galactic centre}
\label{sub:structure_cmz}

The Galactic centre environment has been well studied
\citep{morris_serabyn1996,ferriere2007}. Here we summarise the
structure and physical properties of the molecular gas within a
Galacto-centric radius ($r_{GC}$) of 500\,pc ($|l|<3.5^\circ$) -- ie
the central molecular zone (CMZ). The CMZ can be decomposed into an
outer and inner component, separated at $r_{GC}\sim230$\,pc, and
further separated the inner CMZ into a ``disk'' and torus, with
approximate radial ranges of 0 to 120\,pc and 130 to 230\,pc,
respectively \citep{launhardt2002}. Based on the far-IR dust continuum
emission \citep{launhardt2002}, the mass of the disk and torus
components are derived to be $\sim$4$\times$10$^6$\,M$_\odot$ and
$\sim$1.6$\times$10$^7$\,M$_\odot$, respectively, with a total CMZ
mass ($r_{GC} \leq 500$\,pc) of
$\sim$6$\times$10$^7$\,M$_\odot$. Despite the many systematic
uncertainties in dust-derived masses, and the notoriously uncertain
X-factor affecting CO-derived mass estimates, there is general
agreement in the literature that the total molecular gas mass within
$r_{GC}\leq500$\,pc is $2-6\times$10$^7$\,M$_\odot$
\citep{ferriere2007}.

Numerous studies have argued for two distinct molecular gas
components: a warmer, low-density component ($\leq$10$^3$cm$^{-3}$) of
``diffuse'' clouds with a large filling factor and comprising
$\sim$30\% of the mass, and a high-density ($\geq$10$^4$cm$^{-3}$)
component with a volume filling factor of roughly a few percent
\citep[see][for details]{ferriere2007}. Given the effective critical
density of the $\nhone$, it is likely the HOPS observations are
tracing the dense component.

\begin{figure*}
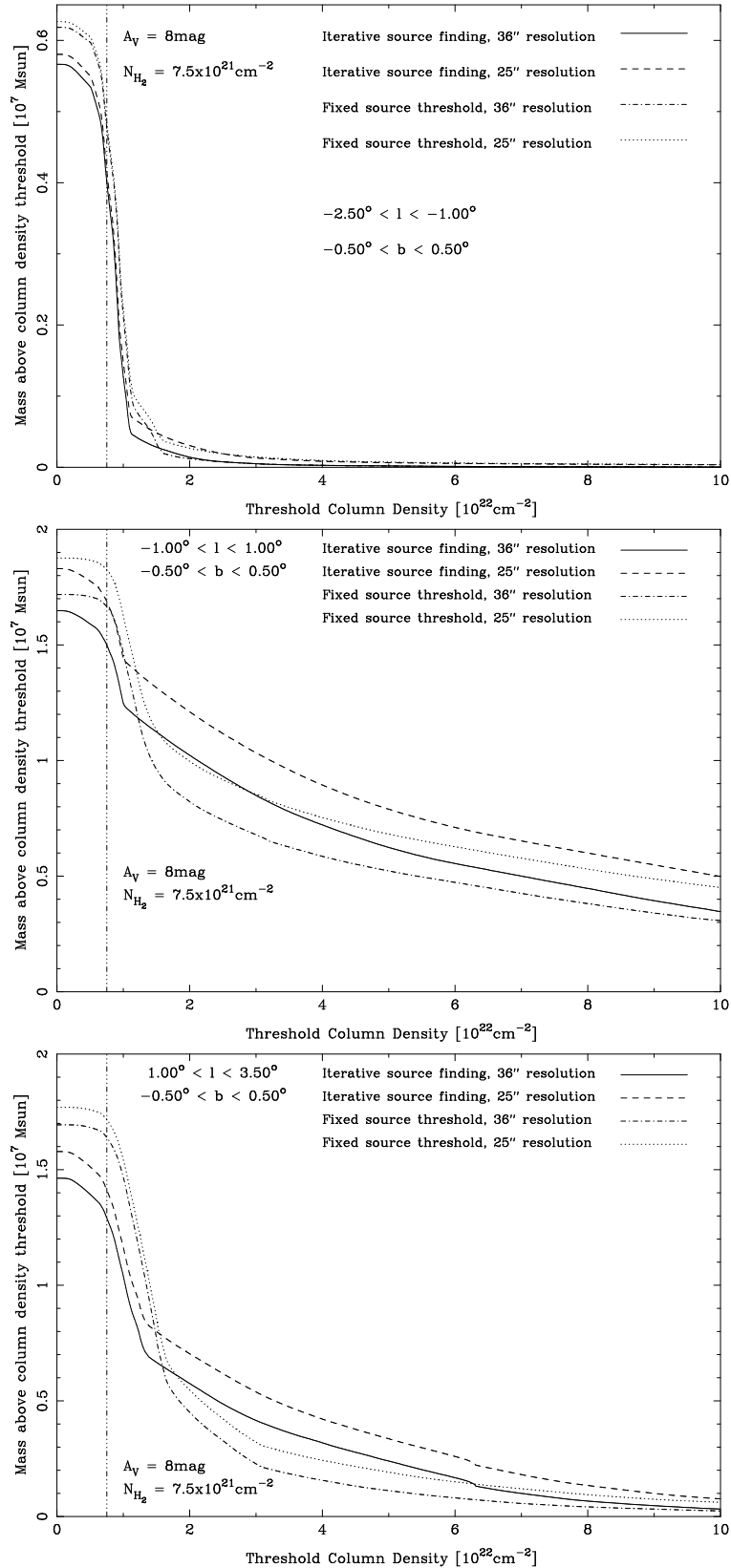

\begin{center}
\includegraphics[height=10cm, angle=-90, trim=0 0 -5 0]{fig4a.ps} \\
\includegraphics[height=10cm, angle=-90, trim=0 0 -5 0]{fig4b.ps} \\
\includegraphics[height=10cm, angle=-90, trim=0 0 -5 0]{fig4c.ps} \\
\end{center}
\caption{\small{Total mass in various longitude ranges of the CMZ
    which lies above a range of column density thresholds. The
    vertical dashed line shows the proposed column density threshold
    of 7.5$\times$10$^{21}$\,cm$^{-2}$, calculated from the extinction
    threshold of $A_V = 8$\,mag and assuming an $A_V \rightarrow
    N_{H_2}$ conversion of $N_{H_2} = A_V \times
    0.95\times10^{21}$\,cm$^{-2}$. The different curves show the
    effect on the derived column density of using different wavelength
    bands and different source-extraction algorithms. From this we
    conclude the column densities are robust to the 10-20\% level.}}
\label{fig:cmz_enc_mass_N_threshold}
\end{figure*}

The recent far-IR HiGAL survey provides a new, high angular resolution
view of the CMZ molecular gas properties \citep{molinari2011}. We
derive the column density maps for the region $-2.5^\circ < l <
3.5^\circ$, $|b|\leq0.5^\circ$ using well-tested methods
\citep{battersby2011}. A major systematic uncertainty is the assumed
dust opacity and we note that we systematically underestimates the
mass by a factor 2$-$3 compared to the DUSTEM method
\citep[e.g.][]{molinari2011} for this
reason. Figure~\ref{fig:cmz_enc_mass_N_threshold} shows the total mass
of gas which lies above a range of different column density
thresholds.  The different curves show the effect on the derived
column density of using different wavelength bands and different
source-extraction algorithms. From this we conclude the mass derived
from the column densities to be robust to the background and source
extraction to the 10-20\% level.

The vertical lines in Figure~\ref{fig:cmz_enc_mass_N_threshold} show
the \citet{lada2012} proposed column density threshold of
7.5$\times$10$^{21}$\,cm$^{-2}$, calculated from the \citet{lada2012}
extinction threshold of $A_V = 8$\,mag and assuming an $A_V
\rightarrow N_{H_2}$ conversion of $N_{H_2} = A_V \times
0.95\times10^{21}$\,cm$^{-2}$ \citep{frerking1982}. This shows that
most of the mass in this region lies above an extinction of $A_V =
8$\,mag. The masses derived from this analysis are reported in
Table~\ref{tab:mass_sfr}. The total derived mass of the CMZ gas from
the HiGAL data is 4.1$\times$10$^7$\,M$_\odot$, in good agreement with
previous measurements in the literature \citep{ferriere2007}.

Determining the volume density of the gas from the HiGAL column
density maps is not straightforward as it depends on knowledge of the
3D gas structure. For the most part this is not well constrained
towards the Galactic centre. The region of the CMZ with the best
quantified 3D structure is the molecular torus, or ``100\,pc
ring'' \citep{molinari2011}. The volume of gas in a ring of radius $R$
and width $\Delta R$, with a thickness of $H$, is
\be  
V = 2\pi R\Delta R H = 4.4\times10^{60}
\left({R\over 100\pc}\right)
\left({\Delta R\over 12\pc}\right)
\left({H\over 20\pc}\right)
{\rm cm}^{-3}
\ee  
\noindent
The fiducial radius of 100\,pc is that determined for the ring and the
fiducial width of 12\,pc is based on the observed column density maps
\citep{molinari2011}. We have used $H=20\pc$ assuming hydrostatic
equilibrium ($H/R=v_T/v_c$); the circular velocity $v_c\approx100\kms$
while the turbulent velocity $v_T\equiv\sqrt{8\ln2}\sigma_v=20\kms$
\citep{oka2001}.

With a mass of 1.8$\times$10$^7$\,M$_\odot$ (see column 5 of
Table~\ref{tab:mass_sfr}), the volume density for the ring is then
\be  
n= {M_g \over m_p V} \approx 5000
\left({R\over 100\pc}\right)^{-1}
\left({\Delta R\over 12\pc}\right)^{-1}
\left({H\over 20\pc}\right)^{-1}
{\rm cm}^{-3}.
\ee  

Note that this mass is a factor $\sim$1.7 smaller than reported
previously \citep{molinari2011} due to the different dust opacities
used in that work. The inferred densities in this work are therefore
also lower by the same factor.  The surface density of the ring is
\be  
\Sigma_{ring}={M_g\over 2\pi R\Delta R}\approx2400 
\left({R\over100\pc}\right)
\left({\Delta R\over12\pc}\right)
M_\odot\pc^{-2},
\ee  
or
\be  
\Sigma_{ring}=0.5\g\,{\rm cm}^{-2}.
\ee  
As the gas in the ring is observed to be clumpy, the density at
individual positions along the ring may clearly deviate from these
representative values.

\begin{figure*}
\begin{center}
\includegraphics[height=15cm, angle=-90, trim=0 0 -5 0]{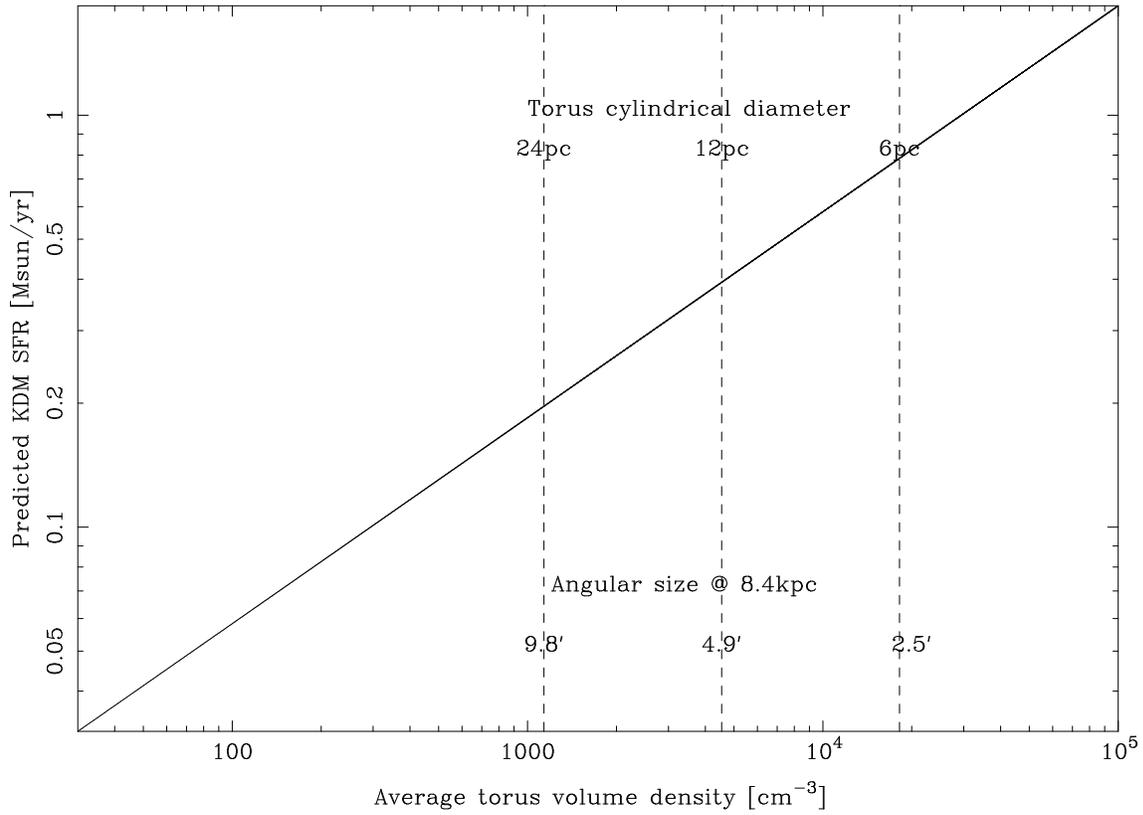} \\
\end{center}
\caption{\small{The predicted star formation rate from the
    \citet{kdm2012} volumetric relation as a function of the average
    gas volume density. The dashed vertical lines show the average
    volume density assuming the $1.8\times10^7$\,M$_\odot$ of gas in
    the region $|l|<1^\circ$, $|b|<0.5^\circ$ (see
    Table~\ref{tab:mass_sfr}) lies in a cylindrical ring with
    semi-major and semi-minor axes of 100\,pc and 60\,pc,
    respectively, with the diameter specified by the annotation for
    each line. The observed thickness (and hence the inferred
    diameter) of the ring varies from $\sim$6\,pc to factors of
    several larger. We estimate the average volume density to be
    $\sim5\times10^3$\,cm$^{-3}$, with a corresponding predicted SFR
    of $\sim$0.4\,M$_\odot$/yr.}}
\label{fig:cmz_density_kdm_sfr}
\end{figure*}

The gas in the ring has been modelled as a twisted ellipse with
semi-major and semi-minor axes of 100\,pc and 60\,pc, respectively
\citep{molinari2011}.  The dashed vertical lines in
Figure~\ref{fig:cmz_density_kdm_sfr} show the average volume density
assuming the ellipse has the average cylindrical diameter (ie
$H=\Delta R$) specified by the annotated values. From the measured
height of the ring projected on the plane of the sky we estimate the
average volume density to be $\sim5\times10^3$\,cm$^{-3}$. This agrees
well with the analytical estimate above and also volume densities
estimated by previous authors for the dense component of the CMZ
molecular gas (see above).

We note that the gas volume density and interpretation of the measured
column density changes considerably if the gas structure is different
than assumed above. For example, in the worst-case scenario, if the
gas assumed to make up the ring actually lay in a uniform disk with
the same radius, the volume density would drop to around a few
$10^2$\,cm$^{-3}$. Additionally, the measured column density would be
highly dependent on the viewing angle, and appear much larger if
viewed through the disk-plane rather than in a face-on view of the
Galaxy. In this case it would not be fair to test the predictions of
the column density threshold SF relations using the directly measured
column density without taking this into account.

However, several lines of evidence suggest this is not the case. On
top of the modelling which successfully reproduces the gas kinematic
structure \citep{molinari2011}, analysis of emission from many high
critical density molecular line transitions show the majority of the
gas in the CMZ is cold, has a density $\gtrsim$10$^4$\,cm$^{-3}$ and a
volume filling factor of only a few percent. In addition, if the gas
was uniformly distributed, one would expect optically thin emission
(e.g. from sub-mm dust emission) to peak along the line of sight with
the largest path length, i.e. directly through the centre. In fact the
emission intensity increases towards the edges, directly contradicting
what would be expected from a uniform disk, and as expected from the
emission arising in a ring.

\subsection{Measured CMZ SFR values}
Values of the SFR for the CMZ in the literature vary by factors of
several, but recent measurements based on number counts of massive
young stellar objects, appear to converge to a value close to
0.1\,$M_\odot {\rm yr}^{-1}$ \citep[e.g. 0.14\,$M_\odot {\rm yr}^{-1}$
  and 0.08\,$M_\odot {\rm yr}^{-1}$:][]{yusef-zadeh2009,immer2012}.

As an alternative approach, we make use of the recent analysis of WMAP
data \citep{murray_rahman2010} to determine the fraction of ionising
photons in the Galaxy coming from the GC. This analysis offers many
advantages over older estimates of the ionising photon flux. WMAP has
much better absolute flux precision, suffers less from spatial
filtering and observed at multiple frequencies so the spectral index
of the emission can be calculated allowing the relative contributions
from free-free, non-thermal and spinning dust emission to be
determined. Although the WMAP beam is large, $\sim1^\circ$ at the
lowest frequencies, the free-free emission in the Galaxy is dominated
by the so-called ``extended low density'' (ELD) \hii emission, rather
than classical \hii regions \citep{murray_rahman2010}.

Based on the free-free-only contribution to the 33\,GHz flux,
corrected for dust attenuation, the rate of ionising photons, $Q$,
across the Galaxy is measured to be 2.9$\times10^{53}$\,$s^{-1}$
\citep{lee2012}. The SFR, $\dot M_*$, is then given by
\citep{murray_rahman2010}

\be \dot M_* = Q{\langle m_*\rangle\over \langle q\rangle} {1\over
  \langle t_Q\rangle}, \ee 

\noindent
where $\langle q\rangle$ is the ionising flux per star averaged over
the stellar initial mass function and $\langle m_*\rangle$ is the mean
mass per star. The quantity $\langle t_Q\rangle$ is the
ionisation-weighted stellar lifetime, or time at which the ionising
flux of a star falls to half its maximum value, averaged over the
IMF\footnote{Assuming high mass stars are well selected from the IMF
  as expected if their formation is intrinsically linked with the
  formation of massive stellar clusters
  \citep[e.g.][]{smith2009}. However, this assumption may be affected
  if stochastic sampling of the IMF is important \citep[see
    e.g.][]{bressert2012a}.}.  \citet{murray_rahman2010} show that
\be  
{\langle m_*\rangle\over \langle q\rangle}=1.59\times10^{-47}M_\odot \s
\ee  
and
\be  
\langle t_Q\rangle = 3.9\times10^6\yr.
\ee  

\noindent
The reason this timescale is small compared to the average lifetime of
stars in a cluster is because $\langle t_Q\rangle$ is dominated by
stars $>$40\,M$_\odot$ \citep{murray2011}. While stars of
$\geq$8\,M$_\odot$ (with lifetimes significantly longer than $\langle
t_Q\rangle$) produce ionising photons, the main-sequence life time of
stars $>$40\,M$_\odot$ is a slowly varying function of mass, while the
ionising photon output continues to increase rapidly. The WMAP
observations are therefore sensitive to star formation over the last
$\sim$4\,Myr.

From the total rate of Lyman continuum photons in the Galaxy of
$Q=$2.9$\times10^{53}$\,$s^{-1}$, the total SFR is calculated to be
1.3\,M$_\odot yr^{-1}$ \citep{murray_rahman2010}. In
Table~\ref{tab:mass_sfr} we show the $Q$ values for the same latitude
and longitude regions as the previous gas mass calculations. The range
of values in column 7 show the total $Q$ assuming the near and far
kinematic distances. Column 9 gives the fraction of the total Galactic
$Q$ that the average of these values corresponds to. Based on this
analysis, for the latitude range of $|b|<0.5^{\circ}$, the longitude
ranges $-2.5^\circ<l<-1.0^\circ$, $|l|<1^\circ$ and
$1^\circ<l<3.5^\circ$ therefore have SFRs of $\sim$0.016, $\sim$0.015
and $\sim$0.0036\,M$_\odot$\,yr$^{-1}$, respectively. We report these
numbers in Table~\ref{tab:mass_sfr}.

This means that in total $\sim$2.8\% of the star formation in the
Galaxy lie in this $l$ and $b$ range. Although this encompasses the
region most people would use to define the CMZ, given the different
angular resolutions of the studies used to derive the mass and SFR, we
check to make sure the hard $l$ and $b$ boundary imposed is not
affecting this result. The WMAP data extend to higher latitudes, and
we note that including the latitude range $|b|<1^\circ$ for the
$|l|<1^\circ$ region increases the maximum $Q$ by a factor
$\sim$3. The total SFR over this larger region is
0.06\,M$_\odot$\,yr$^{-1}$. The $Q$ values over the larger latitude
range are shown in parentheses in columns 7, 9 and 10 of
Table~\ref{tab:mass_sfr}. The other longitude regions are not affected
by increasing the latitude range. As a final check, we note that the
total $Q$ in the much larger region $|l|<4^\circ$, $|b|<1^\circ$ is
1.64$\times10^{52}$, corresponding to 5.6\% of the total Galactic SFR.

\subsection{Comparing measured CMZ gas mass and SFRs with those predicted from SF relations}
We now compare the measured gas mass and SFRs determined for the CMZ
above, with the values predicted from the column-density threshold and
volumetric SF relations.

\subsubsection{Column-density threshold SF relation: predicted CMZ SFR}

The \citet{lada2012} column-density threshold relation proposes the
SFR in a molecular cloud is set by the amount of gas above a column
density threshold corresponding to an extinction of $A_V=8\,$mag. This
is parametrised through, $SFR = 4.6 \times 10^{-8} (f_D)\,
M_{gas}(M_\odot)$\,M$_\odot$\,yr$^{-1}$, where the factor $f_D$ is the
fraction of the total gas mass ($M_{gas}$) above the column density
threshold. Table~\ref{tab:mass_sfr} shows the mass of gas above the
column density threshold for the three different CMZ regions. Based on
these mass measurements, the predicted SFRs for these regions are
0.18, 0.78 and 0.74\,M$_\odot$yr$^{-1}$. These values are listed in
column 11 of Table~\ref{tab:mass_sfr}.

\subsubsection{Volumetric SF relation: predicted CMZ SFR}

Next we consider the \citet{kdm2012} volumetric star formation
relation which is parametrised through:

\be
  \frac{SFR}{[volume]} = \eta \frac{M_{\rm mol}}{[volume]} \,\,\, \frac{1}{\tau_{\rm ff}} = 0.01 \frac{M_{\rm mol}}{[volume]} \,\,\, \frac{1}{\tau_{\rm ff}}
\label{eq:kdm}
\ee

\noindent
where the SFR in a given volume of gas per free-fall time ($\tau_{\rm
  ff}$) is determined by the mass of the molecular gas ($M_{\rm mol}$)
in that volume and a global efficiency of gas converted to stars per
free-fall time, $\eta$, of 1\%. The free-fall time is

\be  
\tau_{ff}\equiv\sqrt{3\pi\over 32 G\rho}=3.6\times10^5\left({10^4\, {\rm cm}^{-3}\over n}\right)^{1/2}\yr.
\label{eq:tff}
\ee  

The free-fall time, and hence gas density, for the gas in the GC
clearly plays an important role in setting the predicted SFR. As
discussed in $\S$~\ref{sub:structure_cmz}, the region of the CMZ with
the best constrained gas structure is the 100\,pc ring (see
$\S$~\ref{sub:structure_cmz}). From Eq~\ref{eq:kdm}~\&~\ref{eq:tff},
the SFR predicted by the volumetric relation is given by,
\be 
\dot M_* = \eta M_g/\tau_{ff} = 0.53
\left({n\over8000\, {\rm cm}^{-3}}\right)^{1/2} M_\odot\yr^{-1}, \ee 

\noindent
for $\eta=0.01$.  This is twenty times the observed star formation
rate. The fiducial predicted SFR increases to 0.9\,M$_\odot\yr^{-1}$
for $\eta=0.017$ \citep[e.g.][]{kennicutt1998}.

Given the clumpy nature of the gas in the ring, for illustrative
purposes, Figure~\ref{fig:cmz_density_kdm_sfr} plots the predicted SFR
as a function of the gas density. The dashed lines show the range in
observed thickness \citep{molinari2011}, or scale height $H$ (and
hence the inferred radial extent $\Delta R$ if the ring is
cylindrical) of the ring from $\sim$6\,pc to factors of several
larger.

We note that, if instead of using the observed geometry of gas, one
assumes the GC gas mass is spread over a uniform disk of diameter
$\sim$100\,pc, the expected free-fall time significantly increases,
and the predicted star formation rate drops considerably. Using this
much larger free-fall time might be appropriate if the timescale for
star formation in the GC was set by the time it would take the gas to
collapse to the centre of the ring. However, the observed gas dynamics
show that the ring is not collapsing towards the centre, but rather
rotating around the centre of the ring \citep{molinari2011}. In
addition the gas in the ring clearly fragments, and the free-fall time
at the observed local density is much shorter that the time it would
take to collapse to the centre (assuming no rotational support). We
conclude that the observed local volume density of
$\sim5-8\times10^3$\,cm$^{-3}$ is the correct density to use for the
purpose of calculating the free-fall time.

We believe the reason our results contradict previous authors who have
concluded the CMZ lies on the S-K relation
\citep[e.g.][]{yusef-zadeh2009,kennicutt_evans2012} is because they
assumed the observed CMZ gas lay in a uniform disk and thus had a much
lower average volume density.

Assuming the rest of the gas in the outer CMZ is at a similar density
to that in the torus \citep{ferriere2007}, the predicted SFR for the
$-3.5^\circ<l<-1^\circ$ and $1^\circ<l<2.5^\circ$ regions are 0.14 and
0.39\,M$_\odot$yr$^{-1}$, respectively.  The predicted SFR values are
listed in column 12 of Table~\ref{tab:mass_sfr}.

\subsubsection{Predicted vs measured SFR values}

The predicted SFRs in Table~\ref{tab:mass_sfr} for both the column
density threshold and volumetric SF relations agree to within a factor
of two. The measured SFRs are at least a factor 10 smaller than the
predicted values for all three regions. The region of the outer CMZ at
positive longitudes (ie $1^\circ<l<3.5^\circ$) stands out as extreme
in this regard. Despite containing almost the same reservoir of gas as
the inner CMZ (ie $|l|<1^\circ$), it has a much lower measured
SFR. Both the column density threshold and volumetric SF relations
over-predict the SFR in this region by two orders of magnitude.

However, the SFR predictions for the volumetric SF relation in
Table~\ref{tab:mass_sfr} rely on the assumption that the volume
density of the outer CMZ is similar to that determined for the inner
CMZ ``100\,pc ring''. Given the uncertainties in determining the gas
density from the measured surface density, we can approach the problem
from a different direction and ask, what gas densities would be
required to produce the measured SFR from the measured gas masses? As
shown in Table~\ref{tab:mass_sfr}, the $1^\circ<l<3.5^\circ$ region
contains 0.85\% of the total molecular gas in the Galaxy and 0.3\% of
the total SF. From Eq~\ref{eq:kdm}, the density of the clouds in this
region should be 0.1 times the density of those in the disk, or
0.1$\times$250\,cm$^{-3}\sim25$\,cm$^{-3}$. This is unfeasibly low.

We conclude the column density and volumetric SF relations over-predict
the SFR in the CMZ by an order of magnitude.

\section{Summary and Discussion}
\label{sec:summary}

In summary, we find that the dense gas ($\nhthree$ \& 500\,$\mu$m) and
SF activity tracers (masers \& HII regions) used in this study are
reliably tracing the present-day relative dense gas mass and SF
activity distributions, respectively, across the Galaxy. We conclude
that the striking difference between the dense gas and SF activity
tracers between the GC-only and non-GC regions shows the current star
formation rate per unit mass of dense gas is an order of magnitude
smaller in the GC than in the rest of the Galaxy. We directly test the
predictions of proposed column-density threshold and volumetric star
formation relations and find, given the mass of dense gas in the GC,
they over-predict the observed SFR in the GC by an order of
magnitude. We conclude the current star formation relations are
incomplete in some way. Any universal column/volume density relations
must be a \emph{necessary but not sufficient} condition for SF to
occur.

Putting the CMZ in the context of the Galaxy as a whole, the Milky Way
contains $\sim2\times10^9\,$M$_\odot$ of gas, so the CMZ holds roughly
a few percent of this. The WMAP analysis shows the CMZ also contains a
few percent of the ionising photons, and hence star formation, in the
Galaxy. Both the volumetric and surface density SF relations predict
that a given mass of gas will form stars more rapidly if the
respective densities are larger. Yet the gas in the CMZ, which has a
much higher surface and volume density than any comparable mass of gas
in the disk of the Milky Way, forms stars at a rate proportional to
the ratio of gas in the CMZ to that in the Milky Way. 

Something is required to slow the rate of SF in the CMZ compared to
that in the rest of the Milky Way. An additional support mechanism,
not taken into account in either the column density threshold or
volumetric SF relations, may be responsible for inhibiting the SF.
One potential solution is therefore an additional term or threshold in
the proposed SF relations. The most noticeable difference between
clouds in the CMZ and the rest of the Milky Way, apart from the 1 to 2
order of magnitude larger volume density, is the order of magnitude
larger internal cloud velocity dispersion
\citep{morris_serabyn1996,ferriere2007}. We thus conclude it is likely
that the relevant term is related to the additional turbulent energy
in the gas providing support against gravitational collapse. A simple
way to parametrise this is through the linewidth ratio, $\Delta V_{\rm
  ratio} = \Delta V/\Delta V_0$, where $\Delta V_0$ is the typical
internal cloud velocity dispersion in disk molecular clouds. The
Schmidt law could then be re-expressed either in terms of surface
density, $\Sigma_{SFR} \propto (\Sigma_{\rm gas})^\alpha/(\Delta
V_{\rm ratio})^b$ [$\alpha \sim$ 1 to 1.4; $b\sim1$], or mass of gas
above the density threshold, $M_{\rm dense}$, through, $\Sigma_{SFR}
\propto M_{\rm dense} /(\Delta V_{\rm ratio})^b$. This would reconcile
the star formation in the CMZ with that in the rest of the
Galaxy. Theorists have wrestled with this before
\citep[e.g.][]{krumholz_mckee2005,padoan_nordlund2011,kdm2012} and we
are currently seeking to test these scenarios.

However, even if the extreme environmental conditions in the CMZ do
inhibit SF, they do not stop it entirely. The CMZ contains Sgr B2, one
of the most extreme cluster forming regions in the Galaxy. It also
contains high-mass star clusters like the Arches and Quintuplet, and
at least one molecular cloud which appears to be the progenitor of
such massive stellar clusters
\citep{longmore2012_brick,bressert2012b}. Evidence exists of episodic
SF events in the CMZ
\citep{sofue_handa1984,yusef-zadeh2009,su2010,bland-hawthorn_cohen2003}
and mechanisms exist to explain how such episodic SF can occur. Gas in
barred spiral galaxies like the Milky Way is funnelled from the disk
through the bar to the GC \citep{kormendy_kennicutt2004,sheth2005}. If
gas is continually fed from the disk to the GC, and the environmental
conditions impose a higher threshold for SF to occur, the gas might
build up until reaching a critical point before undergoing a burst of
SF.

Although clouds near the Galactic centre in the Milky Way and other
galaxies may only represent a small fractional volume of a galaxy,
they can contribute a significant fraction of the total dense
molecular gas. In terms of dense gas mass and environmental
conditions, the Galactic centre also acts as a bridge between local SF
regions in our Galaxy and SF environments in external
galaxies. Understanding why such large reservoirs of dense gas deviate
from commonly assumed SF relations is of fundamental importance and
may help in the quest to understand SF in more extreme (dense)
environments, like those found in interacting galaxies and at earlier
epochs of the Universe.

\bibliography{sfr}


\newpage
\begin{table*}
\begin{center}
  \caption{{\bf Properties of the Galactic plane surveys used in this
    work.  }}
\vspace{1mm}
\label{tab:surveys}
\begin{scriptsize}
\begin{tabular}{|c|c|c|c|c|c|c|c|l|}
\hline \hline
Survey       & Angular          & Transition or             & Frequency or & l$_{min}$    & l$_{max}$   & b$_{min}$     &  b$_{max}$   & Notes\\ 
             & Resolution       & Continuum                 & wavelength   &             &             &              &             &       \\ \hline

GBT HRDS     & 82$\arcsec$      & H86$\alpha$$-$H93$\alpha$ & 9\,GHz       & -17$^\circ$  & 67$^\circ$  & -1$^\circ$    & 1$^\circ$    & Targetted \hii\ region survey\\
HOPS         & 2$\arcmin$       & $\nhone$                  & 23.4\,GHz    & -70$^\circ$  & 30 $^\circ$ & -0.5$^\circ$  & 0.5 $^\circ$ & n$_{crit}\sim10^{3\rightarrow4}$\,cm$^{-3}$\\
HOPS         & 2$\arcmin$       & Water maser               & 22\,GHz      & -70$^\circ$  & 30 $^\circ$ & -0.5$^\circ$  & 0.5 $^\circ$ & Traces SF activity\\
HOPS         & 2$\arcmin$       & H68$\alpha$               & 22\,GHz      & -70$^\circ$  & 30 $^\circ$ & -0.5$^\circ$  & 0.5 $^\circ$ & Traces HII regions\\
MMB          & $\sim$1$\arcsec$ & Methanol  maser           & 6.7\,GHz     & -180$^\circ$  & 20 $^\circ$ & -2$^\circ$    & 2$^\circ$    & Traces SF activity\\
Hi-GAL       & 35$\arcsec$      & continuum                 & 70,160,250,  & -60$^\circ$  & 60$^\circ$  & -1$^\circ$    & 1$^\circ$    & optically-thin dust emission \\ 
             &                  &                           & 350,500\,$\mu$m &          &             &              &             &  (S$_\nu\propto$ M$^{dense}_{gas}$) \\ \hline

\end{tabular}
\end{scriptsize}
\end{center}
\end{table*}


\begin{table*}
\begin{center}
  \caption{Gas mass and ionising photon rate (Q) for various longitude
    and latitude ranges of the CMZ. The second row corresponds roughly
    to the longitude range for the inner $\sim$150\,pc of the CMZ. The
    first and third rows correspond roughly to the longitude ranges of
    the outer $\sim$150 to $\sim$500\,pc of the CMZ, in the negative
    and positive longitude ranges, respectively. The gas masses are
    derived from the HiGAL column density maps. The value in column 5
    is the total mass in that longitude and latitude range, whereas
    column 6 reports the mass above the column density threshold
    corresponding to an extinction of A$_V=8$\,mag (see text for
    details). The range in values in column 7 corresponds to the total
    Q assuming near and far kinematic distances, respectively
    \citep{lee2012}. Column 8 gives the fraction (in percent) of the
    total Galactic molecular gas contained in column 5, assuming a
    total Galactic molecular gas mass of
    2$\times$10$^9$M$_\odot$. Column 9 gives the fraction (in percent)
    of the total Galactic Q of 2.9$\times$10$^{53}$\,s$^{-1}$
    \citep{lee2012}. Column 10 gives the corresponding SFRs for each
    of the regions. The values in parentheses in columns 7, 9 and 10
    are the $Q$ values over the same longitude regions if the latitude
    range is doubled to $|b|<1^\circ$. Only the $|l|<1^\circ$
    longitude range is affected. The SFRs in columns 11 and 12 are
    those predicted from the column density threshold and volumetric
    SF relations, respectively (see text for details).}
  \vspace{1mm}
\label{tab:mass_sfr}
\begin{scriptsize}
\begin{tabular}{|c|c|c|c|c|c|c|c|c|c|c|c|} \hline
$b_{\rm min}$ & $b_{\rm max}$ & $l_{\rm min}$ & $l_{\rm max}$ & mass         & mass$_{\rm Av\geq8}$ & Q                    & mass$/$M$_{TOT}$ & Q/Q$_{TOT}$ & SFR$_Q$                & SFR$_{\rm Av\geq8}$ & SFR$_{\rho}$\\
 (deg)       &  (deg)      &  (deg)       & (deg)    & (10$^7$M$_\odot$) & (10$^7$M$_\odot$)   & (10$^{51}$\,s$^{-1}$) &  (\%)           &  (\%)      &($M_\odot {\rm yr}^{-1}$) &($M_\odot {\rm yr}^{-1}$) &($M_\odot {\rm yr}^{-1}$) \\ \hline

$-0.5$       & 0.5         &  $-2.5$      &  $-1.0$  & 0.6              & 0.4                &  2$-$6               &  0.3            &  1.3       & 0.016             & 0.184 & 0.14\\
$-0.5$       & 0.5         &  $-1$        &  1       & 1.8              & 1.7                &  3$-$4(13)           &  0.9            &  1.2(4.5)  & 0.012-0.018(0.06) & 0.782 & 0.41\\
$-0.5$       & 0.5         &   1          & 3.5      & 1.7              & 1.6                &  0.1$-$2             &  0.85           &  0.3       & 0.0036            & 0.736 & 0.39\\
\hline
\end{tabular}
\end{scriptsize}
\end{center}
\end{table*}


\end{document}